# Reversible hydrogenation and band gap opening of graphene and graphite surfaces probed by scanning tunneling spectroscopy

*Andres Castellanos-Gomez* [1,2,*,+], *Magdalena Wojtaszek* [1], *Arramel* [1], *Nikolaos Tombros* [1] and *Bart. J. van Wees* [1,*]

[1] Physics of Nanodevices, Zernike Institute for Advanced Materials, University of Groningen, The Netherlands.

[2] Departamento de Física de la Materia Condensada (C–III). Universidad Autónoma de Madrid, Campus de Cantoblanco, 28049 Madrid, Spain.

* e-mail: a.castellanosgomez@tudelft.nl , b.j.van.wees@rug.nl
+ Present address: Kavli Institute of Nanoscience, Delft University of Technology, Lorentzweg 1, 2628 CJ Delft (The Netherlands)

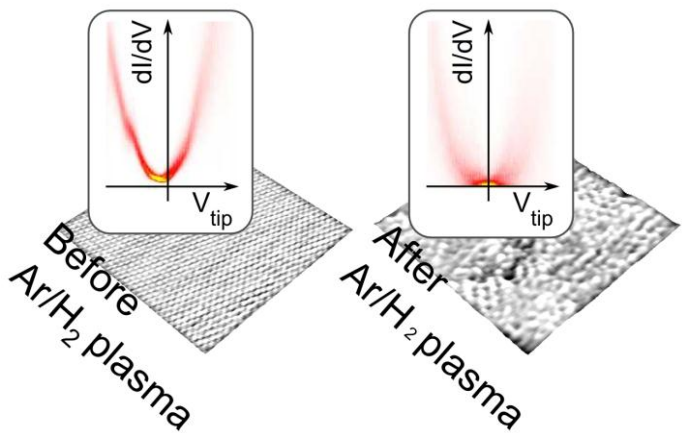

*The effect of hydrogenation on graphene and graphite surfaces is studied by scanning tunneling microscopy and spectroscopy. We employ Ar/H$_2$ plasma to chemically modify the surface of the samples, opening an energy band gap of 0.4 eV. A moderate annealing is enough to close this band gap and the samples can be hydrogenated again yielding a similar semiconducting behavior.*

The effect of hydrogenation on the topography and the electronic properties of graphene and graphite surfaces are studied by scanning tunneling microscopy and spectroscopy. The surfaces are chemically modified using Ar/H$_2$ plasma. Analyzing thousands of scanning tunneling spectroscopy measurements we determine that the hydrogen chemisorption on the surface of graphite/graphene opens on average an energy band gap of 0.4 eV around the Fermi level. We find that although the plasma treatment modifies the surface topography in a non-reversible way, the change in the electronic properties can be reversed by a moderate thermal annealing and the samples can be hydrogenated again yielding a similar, but slightly reduced, semiconducting behavior after the second hydrogenation.

**1. Introduction**

Since its first experimental realization by mechanical exfoliation of graphite on $SiO_2$ surfaces,[1] graphene has attracted a lot of attention because of its unique electronic properties.[2-6] Graphene is a zero gap semiconductor with a linear energy-momentum dispersion relation, which implies that the charge carriers behave as massless Dirac fermions.[7] This fact made possible to experimentally observe very unusual effects such as the fractional quantum Hall effect,[8] the Klein tunneling,[9] and bipolar supercurrent.[10] However, the application of graphene in microelectronic devices is hampered by the lack of a bandgap around the Fermi level. Such an energy bandgap is mandatory in order to fabricate electrically switchable devices with large on-off ratio based on graphene. Up to date several strategies have been developed to open a band gap in graphene. Among them, chemical functionalization of the graphene surface can be very attractive for industrial applications because it is compatible with large scale production of semiconducting graphene. In particular, the exposure of the graphene surface to hydrogen plasma [11-13] or hot atomic hydrogen [14, 15] leads to the chemisorption of hydrogen atoms, which produces the $sp^3$ hybridization of carbon network, reduces the number of delocalized $sp^2$ electrons and consequently opens a band gap. The value of the band gap $E$ depends on the level of hydrogenation, for instance for graphone, [16] a graphene layer with only one side of the surface fully hydrogenated, the calculated value is $E = 0.45$ eV and it is respectively lower for partially hydrogenated graphene surfaces.[17] This property provides a degree of freedom to tailor the graphene transport properties according to the required semiconducting behavior.

In this article we have studied the effect of hydrogenation on graphene by its exposure to hydrogen plasma in reactive ion etching system (RIE), as described in Ref. [18]. Hydrogenation using plasma, next to hot atomic hydrogen exposure, is the most common technique to induce H chemisorbtion in graphene. Its advantages are: high reactivity of incident hydrogen ions and compatibility with standard wafer-scale microfabrication techniques. The acceleration voltage used for feeding the plasma, when too high, can cause graphene etching and its irreversible damage. Although in Ref. [18] it was showed through electronic transport measurements that at properly chosen plasma conditions there is no sputtering of graphene, the microscopic confirmation of this hydrogen plasma effect on graphene is missing so far. We have studied the topographic and electronic changes produced by the chemisorption of hydrogen on top of graphene and graphite surfaces by means of scanning tunneling microscopy and spectroscopy. From a statistical analysis of thousands of STS spectra, acquired at 50-100 different positions on the surface of both materials, we have determined that chemisorption of hydrogen induces the opening of an energy band gap of 0.4 eV around the Fermi level. We have also found that a moderate thermal annealing of the crystals is enough to close this band gap and more interestingly, the samples can be hydrogenated again yielding a similar semiconducting behavior.

## 2. Results

### 2.1. Topographic changes due to the hydrogenation

We have first studied the structural changes of the topography induced by the hydrogenation of the surface with an Ar/$H_2$ plasma treatment (see Ref. [18] and experimental section for more details). Although the presence of Ar gas in the mixture is not strictly necessary for the hydrogenation process, it has been used due to safety considerations. Both highly oriented pyrolitic graphite (HOPG) and graphene/few-layer graphene grown on nickel by chemical vapour deposition (CVD) has been studied. Graphite samples have been cleaved before starting the study while CVD graphene samples have been used as received.[19] According to the manufacturer, the CVD graphene samples present patches 3-10 μm in size with a thickness that ranges from 1 to 4 layers.

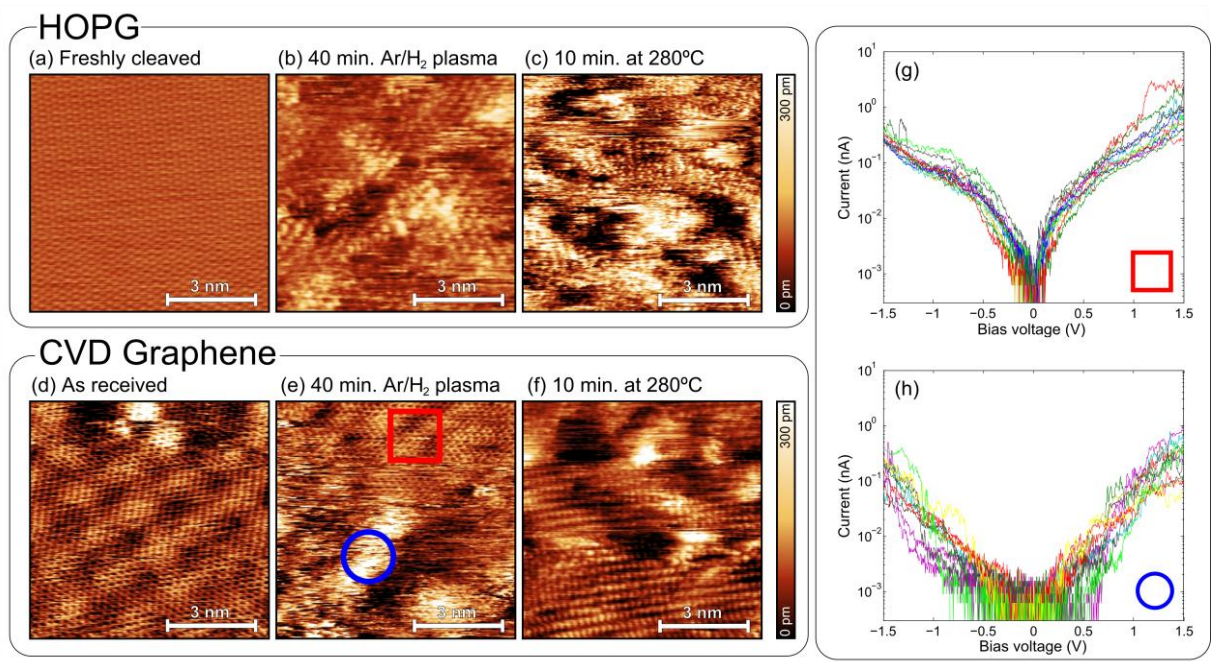

**Figure 1.** Topography images, acquired in the constant current STM mode, (a-c) HOPG graphite (d-f) graphene grown by CVD on top of a nickel surface at different steps of the hydrogenation/dehydrogenation process. (a) and (d) the topography of the surface before the hydrogen plasma treatment. For the HOPG the typical triangular lattice can be resolved all over the surface. For the CVD graphene, a Moiré pattern, due to the lattice mismatch between the graphene and the nickel lattices, is superimposed onto the honeycomb lattice is observed. In (b) and (e), after 40 min. of $H_2$/Ar plasma treatment, the roughness of the surfaces increases. The surfaces are covered with bright spots in where the atomic resolution is lost or strongly distorted. (c) and (f) graphene surface after 10 min. of moderate annealing, the topography of both the HOPG and CVD graphene surfaces do not fully recover their original crystallinity. (g) Current *vs*. voltage traces measured for a CVD graphene sample in several regions with pristine atomic resolution like the one marked with the red square in (e). (h) Same as (g) but measured in several bright regions, like the one marked with the blue circle in (e), where the atomic resolution is distorted.

**Figure 1** shows $8 \times 8$ nm$^2$ topographic images acquired in the constant current STM mode of the surface of HOPG graphite and graphene grown on nickel by CVD. Before the plasma treatment, both samples exhibit highly crystalline surface. In the case of graphite, the typical triangular lattice can be resolved in most of the surface regions (Figure 1(a)).[20] For the CVD graphene one can see a Moiré pattern superimposed on the honeycomb lattice of graphene (Figure 1(d)).[21-27] This Moiré pattern, which is the most commonly observed in our samples, corresponds to the one reported for a graphene monolayer grown onto a nickel (111) surface and it is due to the mismatch between the graphene and the nickel lattices.[28] During the experiments reported here we have studied several tens of locations on the CVD graphene samples, observing occasionally other Moiré patterns associated to few-layer graphene. Therefore, in the CVD graphene samples we sometimes probed regions with 2-4 layers in thickness.

Both graphite and CVD graphene samples are then exposed for 40 minutes to an Ar/H$_2$ plasma to hydrogenate their surfaces. Previous electronic transport measurements indicate the large increase of scattering cross-section for exposure times >1h both in single and bilayer graphene,[18] suggesting the coalescence of hydrogen defects. Moreover Raman spectroscopy measurements have shown that 40 minutes plasma treatment does not introduce a noticeable amount of defects on graphene.[18] Therefore we have chosen the exposure time of 40 minutes, yielding moderate hydrogen coverage, allowing to locally resolve the underneath graphene surface. The hydrogen coverage has been estimated from the ratio between the regions showing the pristine atomic resolution and the bright regions in where the atomic resolution is strongly distorted due to chemisorbed hydrogen. From the analysis of tens of STM topography images, acquired at different locations on the sample, we estimate that the hydrogen absorption modifies about a 30-40% of the surface of the sample both for graphite and CVD grown graphene samples.

The expected structural changes due to hydrogenation are twofold. First, the chemisorption of hydrogen atoms will change the sp$^2$ hybridization of carbon atoms to tetragonal sp$^3$ hybridization, modifying the surface geometry.[29, 30] Second, the impacts of heavy Ar ions, present in the plasma, could also modify the surface by inducing geometrical displacement of carbon atoms (rippling graphene surface) or creating vacancies and other defects. Figure 1(b) and (e) show the topography images of the surfaces of graphite and CVD graphene after the extended plasma treatment. The corrugation increases for both of them after the treatment and there are brighter regions in where the atomic resolution is lost or strongly distorted. This increase of corrugation can be explained by the change of hybridization induced by the chemisorption of hydrogen. [29, 30] We have also found that these bright regions present a semiconducting behavior while the rest of the surface remains conducting (see Figure 1(g)-(h)). The room temperature thermal drift, however, makes it challenging to spatially resolve the electronic properties of the samples by STS with atomic accuracy and thus in the presented approach we focus on statistical properties of the graphene surface. Both the strong distortion of the pristine atomic resolution and the semiconducting behavior can be explained by the accumulation of hydrogen atoms in these bright regions forming clusters. For the CVD graphene samples, the graphene-substrate interaction may play an important role on the spatial distribution of the hydrogen chemisorption for low coverages.[14] Nevertheless, we have not observed any preferential distribution of hydrogen in the CVD graphene samples which is in agreement with a previous work in graphene on SiC in where it was reported that for large hydrogen coverage the chemisorption does not show any preferential position on the graphene lattice.[15]

To confirm whether the hydrogenation or the Ar ion impacts is the main source of these structural changes, we have annealed our samples at moderate temperature. Previous work in similar graphene-based systems [31, 32] indicate that annealing the samples for 10 minutes at 280ºC largely removes the chemisorbed hydrogen from the surface, while it cannot cure possible graphene voids. It is thus interesting to study also the topographic changes produced after the annealing. As shown in Figure 1(c) and (f), the topography of the samples after annealing is similar to the one after the plasma treatment. There are regions in where the graphite/graphene lattice can be well-resolved but there are also few brighter regions indicating that the recovery was not complete. However, after the thermal treatment the brighter regions show horizontal lines and elongated features in the STM topography images which are typical of samples with movable atoms on the surface. These movable atoms can be carbon atoms or atmospheric adsorbates on the surface. As the samples have been annealed, we rule out that these movable atoms are remaining hydrogen atoms in the surface. We have also observed that the STS spectra all over the surface are rather homogeneous, showing a marked conducting behavior.

**2.2. Study of the electronic changes due to the hydrogenation**

It is interesting to study the changes in the electronic properties of both graphite and graphene after the hydrogen plasma treatment to verify if a band gap is opened. In a previous work, electronic transport measurements showed a dramatic increase in the resistance of graphene after hydrogenation.[18] The absence of a systematic study as a function of the temperature, however, hampered the estimation of the bandgap in those measurements. Here we use STS, which is well-known as an extraordinary sensitive technique, to probe the electronic properties of the samples before and after the $H_2$/Ar plasma treatments.

The measured STS spectra, however, typically depend on the exact atomic arrangement of the atoms involved in the tunneling process. Therefore, in ambient conditions the atomic diffusion and the thermal drift yield fluctuations between different measured spectra. To overcome this trace to trace variations we have introduced a procedure to statistically analyze thousands of STS spectra acquired at several tens of different, randomly selected spots in the sample for each step of the hydrogenation/dehydrogenation cycle (see experimental section for a more detailed description of the STS measurement procedure).

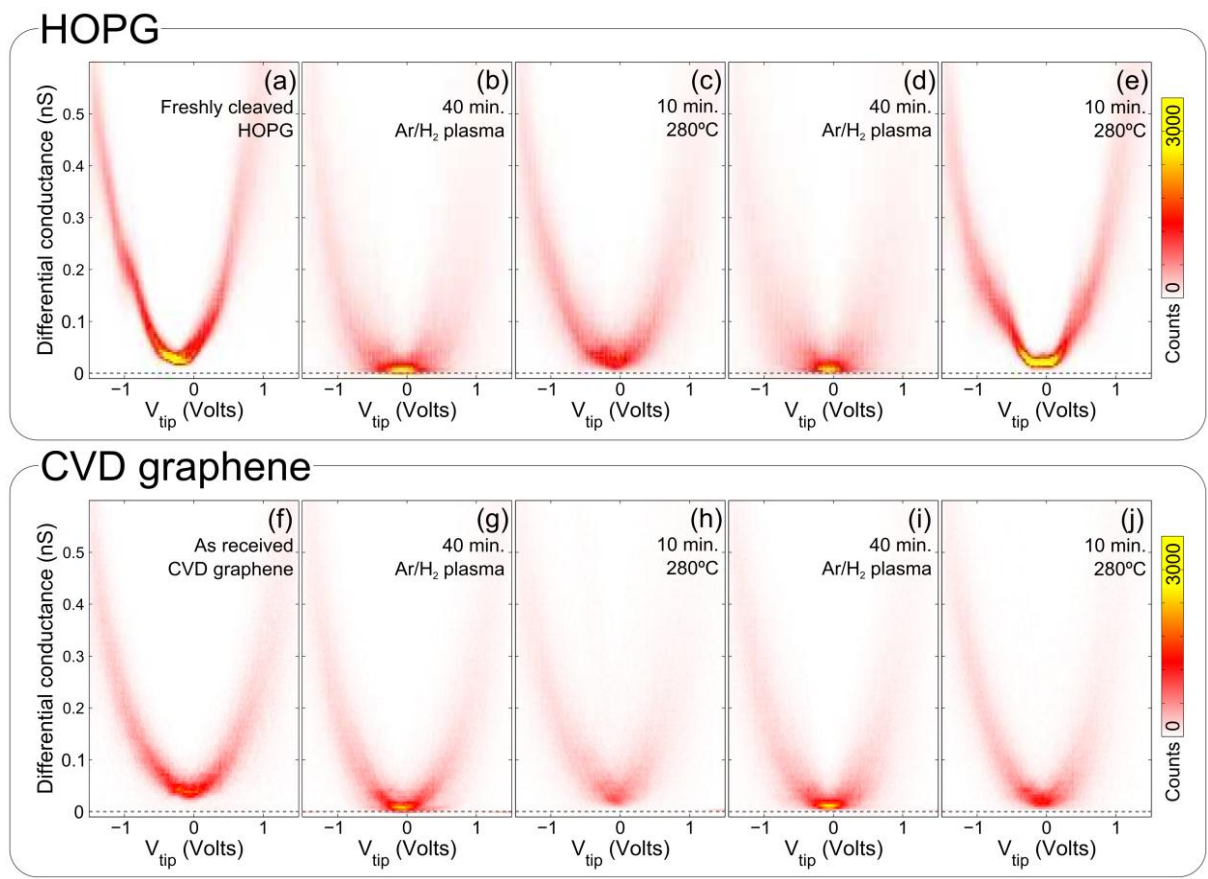

**Figure 2.** Two dimensional histograms of the differential conductance as a function of the tip bias voltage measured for HOPG (a-e) and CVD graphene (f-j) at different steps of the hydrogenation/dehydrogenation process. The samples are hydrogenated/dehydrogenated several times and the changes on their electronic properties are probed after each process step, measuring 2000 STS spectra at 50-100 different positions on the surface. The hydrogenation step is carried out by a 40 min. exposure to an $Ar/H_2$ plasma and the dehydrogenation is done by 10 min. annealing at 280ºC. For both the HOPG and CVD graphene samples we have observed that after the $Ar/H_2$ treatment, the 2D histograms show a semiconducting behavior (accumulation of data points around zero differential conductance value). This behavior disappears after the annealing of the samples, recovering a metallic behavior.

The data corresponding to each step of the hydrogenation/dehydrogenation cycle are represented together, without any selection of the traces, in a two-dimensional histogram (**Figure 2**). This kind of representation has been previously used to determine the most probable STS spectra of a single molecule in a break junction experiment [33] where the configuration of the molecule can change during the experiment. We can use this method to determine the most probable differential conductance vs. voltage (d$I$/d$V$ *vs.* $V$ hereafter) trace on the surface after each step of the plasma/annealing treatment. A general feature of the measured d$I$/d$V$ *vs*. $V$ curves is their parabolic shape for voltages larger than 0.5-0.6 V. This parabolic shape of the tunneling differential conductance can be accounted for with the Simmons model when the tip-sample voltage is comparable to the apparent tunneling barrier height. At low tip-sample voltages, however, the tunneling differential conductance is proportional to the local density of states, giving information about the electronic properties of the sample. For instance, for freshly cleaved graphite (Figure 2a) the d$I$/d$V$ *vs*. $V$ has a non-zero minimum value of 35 ± 11 pS, while the value for the CVD graphene sample is 45 ± 10 pS. The high value of differential conductance measured in CVD graphene samples is in agreement with the metallic behavior observed by Murata et al. in graphene over nickel.[28] Additionally, in Ref. [28] they perform DFT demonstrating that, despite the bandgap opening at the K point of the Brillouin zone,[34] the hybridization of the Ni and C orbitals renders the graphene metallic, explaining the observed increase of the differential conductance at low bias with respect to the graphite.[28] After the plasma treatment, this d$I$/d$V$ *vs*. $V$ presents a clear accumulation of points with zero differential conductance (Figure 2b) as expected for a semiconducting material. After moderate annealing, the samples recover their original conducting behaviour, the differential conductance minimum value for the graphite sample is 31 ± 15 pS and for the CVD graphene 44 ± 20 pS. It is important to note, however, that the increased corrugation of the surface after these treatments causes an inhomogeneous distribution of the electronic properties on the sample which shows up as more blurry histograms (see Figure 2c). After that, the sample is treated again with hydrogen/argon plasma and the semiconducting behavior is again observed (Figure 2d). A final annealing has been used to check that the sample recovers the metallic behavior (Figure 2e).

The case of CVD graphene is remarkably similar to that of graphite. One main difference, however, can be pointed out. After the second plasma treatment the semiconducting behavior of the CVD graphene is less marked, with lower density of zero d$I$/d$V$ counts in STS 2D histogram. Based on our STM images and previous Raman spectroscopy measurements in similar systems we rule out the possible breaking of the graphene layer, exposing the metallic nickel surface, during the plasma treatment. Therefore, we attribute this reduction of the semiconducting behaviour to a modification of the graphene/nickel coupling produced by these plasma/annealing/plasma treatments.

We have additionally studied the role of the Ar ion impacts (occurring during the Ar/H$_2$ plasma treatment) in these changes of the electronic properties. We have found that after a pure Ar plasma treatment the samples do not present the semiconducting behavior observed in samples treated with Ar/H$_2$ plasma (see supporting information).

It is important to note that these two-dimensional d$I$/d$V$ *vs*. $V$ histograms are built with traces measured at different locations in the sample (including both regions highly covered by hydrogen and pristine regions) which show very different electronic behaviour, as shown in Figure 1(g)-(h). Therefore, in order to quantify the appearance of some semiconducting behavior after the hydrogen plasma treatments, we have studied the accumulation of data points around zero d*I*/d*V* in the two-dimensional histograms shown in Figure 2. A line profile along d*I*/d*V* = 0 (dashed lines in Figure 2) is taken to represent the number of counts as a function of the tip voltage (Figure 3). For a conducting sample one would expect a complete absence of counts at d*I*/d*V* = 0. On the other hand, for a semiconducting sample with a well-defined bandgap value, the line profile along d*I*/d*V* = 0 would show a constant number of counts between two voltage values that defines the gap. If one now considers the scenario of a heterogeneous semiconducting sample in which there are spots more conducting than others, the line profile along d*I*/d*V* = 0 would show a broad distribution whose full-width-half-maximum (FWHM) indicates the most probable bandgap on the sample. For both graphite and graphene pristine or annealed samples, these one-dimensional histograms show a negligible number of data points with zero d*I*/d*V* confirming the semi-metallic properties. On the other hand, samples treated with plasma show a strong number of counts with zero d*I*/d*V* which follows a nearly Gaussian distribution as a function of tip-sample bias. The number of zero d*I*/d*V* counts after the second plasma treatment is smaller, especially for the case of CVD graphene.

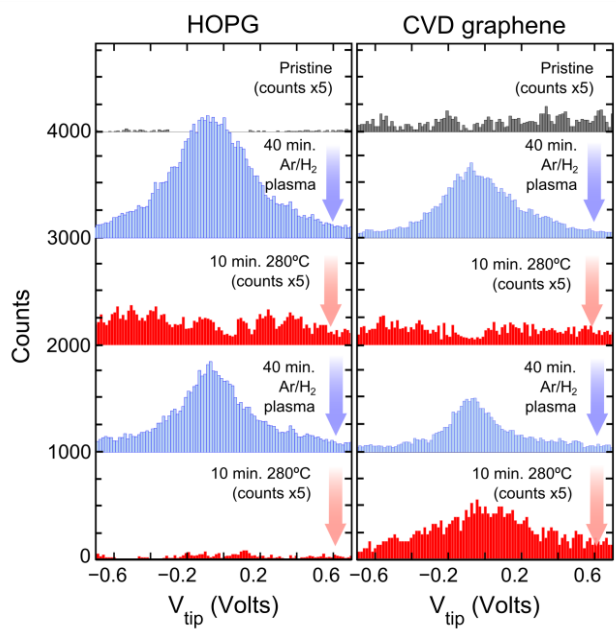

**Figure 3.** One dimensional histograms extracted from a profile along the dashed line (zero differential conductance) in the two dimensional histograms of Figure 2. Left panel corresponds to the HOPG and right panel to CVD graphene sample. The consecutive histograms at different steps of the hydrogenation/dehydrogenation treatments are plotted in the same panel but vertically displaced for clarity. From these histograms one can observe that the number of counts at zero differential conductance for the pristine and annealed samples is negligible in comparison with the number of counts after the Ar/H$_2$ plasma treatments. Note that the counts of the histograms for pristine and annealed samples have been multiplied by 5 to facilitate their comparison.

One can additionally estimate the average energy band gap opened after the plasma treatment from the full width at half maximum of the Gaussian peak in the histograms obtained for the plasma treated samples (**Figure 4**). Surprisingly we have found that even though the topographic changes induced during these treatments are not fully reversible, the opening/closing of the energy band gap is reversible. Moreover, in the case of the HOPG the values of the mean energy band gap obtained after the hydrogenation and after the hydrogenation/dehydrogenation/hydrogenation process are very similar, 0.49 ± 0.04 eV and 0.40 ± 0.04 eV respectively, which is in agreement with the values obtained for partially hydrogenated graphene by other methods.[35] In the case of CVD graphene, the hydrogenated samples show an average band gap opening of 0.45 ± 0.05 eV but the hydrogenated/dehydrogenated/hydrogenated samples show the opening of a band gap with a lower value (0.30 ± 0.03 eV). This reduction in the energy band gap can be due to the observed non-reversible topographic changes that can modify the reactivity of the surface.[36]

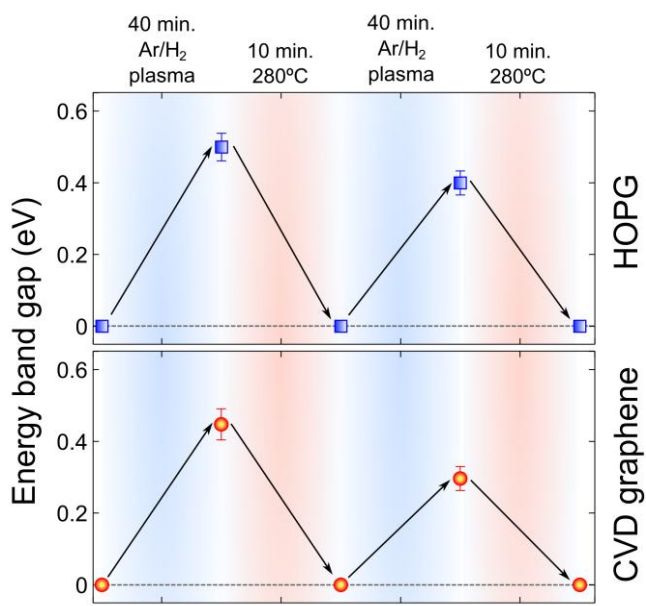

**Figure 4.** Average energy band gap after several hydrogenation/dehydrogenation steps for both HOPG (top panel) and CVD graphene samples (bottom panel). This energy band gap has been determined from the full-width-at-half-maximum of the counts distribution in Figure 3.

3. Conclusions

Using scanning tunneling microscopy and spectroscopy we have studied the effect of hydrogenation on the topography and the electronic properties of graphene and graphite surfaces. An Ar/$H_2$ plasma has been employed to chemically modify the surface of the samples. We have introduced a method to statistically analyze thousands of scanning spectroscopic spectra which made us possible to determine that chemisorption of hydrogen induces the opening of an average energy band gap of 0.4 eV around the Fermi level. Interestingly, although the topographic changes induced by the hydrogenation are not fully-reversible, a moderate annealing of the crystals is enough to close this band gap and the samples can be hydrogenated again yielding a similar semiconducting behavior. The method presented here to average STS data provides a reliable way to study the effect of surface modification in samples where other techniques (such as angle resolved photoemission spectroscopy, Raman spectroscopy, electronic transport measurements) cannot be applied because of, for example, the presence of a conducting substrate underneath.

## 4. Experimental Section

*Graphene and graphite hydrogenation*

All the hydrogenation steps are done using Ar:$H_2$ plasma (composition of 85:15) in reactive ion etching system with a high frequency generator operating at 13.56 MHz, capacitively coupled to the bottom electrode. The gas flow is kept constant at 200 sccm and the pressure in the chamber is 0.05 mbar. We chose the lowest plasma ignition power, P = 3 W (power density is ~ 4 mW/cm$^2$), and tuned the circuit impedance to reduce the built-in DC self-bias between the bottom electrode and the plasma down to zero. As proved in Ref. [18], at the chosen conditions the sputtering of carbon from graphene surface for the samples which are in electrical contact with chamber electrode (the case here) is completely supressed. Each exposure is done for 40 min, which leads to moderate hydrogen coverage and is directly followed by STM/STS measurements. Thermal annealing steps are performed on the hot-plate at 280ºC in the nitrogen environment for 10 min. each.

*Scanning Tunneling Microscopy (STM)*

STM measurements were carried out by utilizing a *PicoLE STM from Agilent Technologies*. The STM tips were obtained by mechanically cutting a high purity $Pt_{0.8}Ir_{0.2}$ wire 0.25 mm in diameter (Goodfellow). The STM images were acquired in constant current operating mode in room conditions. Typical scanning parameters for obtaining STM images of HOPG and CVD-graphene surfaces are in the range between 0.5 to 2 nA and tip bias voltage -0.2 to -0.5 V.

*Scanning Tunneling Spectroscopy (STS)*

The STS measurements are carried out as follows. 50 current *vs.* voltage traces (*IV* traces) are measured at a certain spot on the sample (interrupting the feedback control loop during the measurements). The variation of the tunneling current is measured as the tip bias voltage is swept (0.01 second per trace). By measuring both the forwards and backwards voltage sweep we checked that the thermal drift during the acquisition of a single *IV* trace can be neglected and thus every *IV* trace is measured at a fixed tip location. Then we change the lateral position of the tip by 50-100 nm and another set of 50 traces are acquired. It is important to note that the spectra have been collected without distinction between the bright regions or the regions where the atomic resolution is retained. The procedure is repeated until 2000 traces are collected. The differential conductance *vs.* voltage (d*I*/d*V vs. V*) is obtained by numerical differentiation of the *IV* traces. Then the whole set of 2000 d*I*/d*V vs. V* are represented together in a 2D histogram which shows the most probable shape of the d*I*/d*V vs. V* traces on the sample. It is important to note that these two-dimensional d*I*/d*V vs. V* histograms are built with traces measured at different locations in the sample (including both regions highly covered by hydrogen and pristine regions) which will show very different electronic behaviour, as shown in Figure 1(g)-(h). To build these 2D histograms both the bias voltage and the d*I*/d*V* axes are discretized into *N* number of bins forming an *N* by *N* matrix (200 by 200 in our case). Each datapoint whose d*I*/d*V* and *V* values are within the interval of one bin, adds one count to it. The number of counts in each bin is then represented with a color scale.

**Acknowledgements**

A.C-G. acknowledges fellowship support from the Comunidad de Madrid (Spain) and the Universidad Autonoma de Madrid (Spain). M.W., and N.T. acknowledge financial support from the Ubbo Emmius program of the Groningen Graduate School of Science, the Zernike Institute for Advanced Materials and the Netherlands Organization for Scientific Research (NWO-CW) through a VENI grant. A. thanked financial support from the Foundation for Fundamental Research on Matter (FOM G-08).

# Supplementary information:

# Reversible hydrogenation of graphene and graphite probed by scanning tunneling spectroscopy

*Andres Castellanos-Gomez* [1,2,*,+], *Magdalena Wojtaszek* [1], *. Arramel* [1], *Nikolaos Tombros* [1] *and Bart. J. van Wees* [1,*]

[1] Physics of Nanodevices, Zernike Institute for Advanced Materials, University of Groningen, The Netherlands.
[2] Departamento de Física de la Materia Condensada (C–III). Universidad Autónoma de Madrid, Campus de Cantoblanco, 28049 Madrid, Spain.

* e-mail: a.castellanosgomez@tudelft.nl , b.j.van.wees@rug.nl
[+] Present address: Kavli Institute of Nanoscience, Delft University of Technology, Lorentzweg 1, 2628 CJ Delft (The Netherlands)

## Argon plasma treatment

We have studied the role of the Ar ions impact (occurring during the Ar/$H_2$ plasma treatment) in the changes of the electronic properties of graphite and graphene. The samples are exposed for 40 min. to pure Ar plasma. After the treatment we perform a statistical analysis of 2000 scanning tunneling spectra acquired at 50-100 different spots in the samples surfaces. We have found that after a pure Ar plasma treatment the samples do not present the semiconducting behavior (accumulation of counts in the zero differential conductance region of the 2D histograms) observed in samples treated with Ar/$H_2$ plasma.

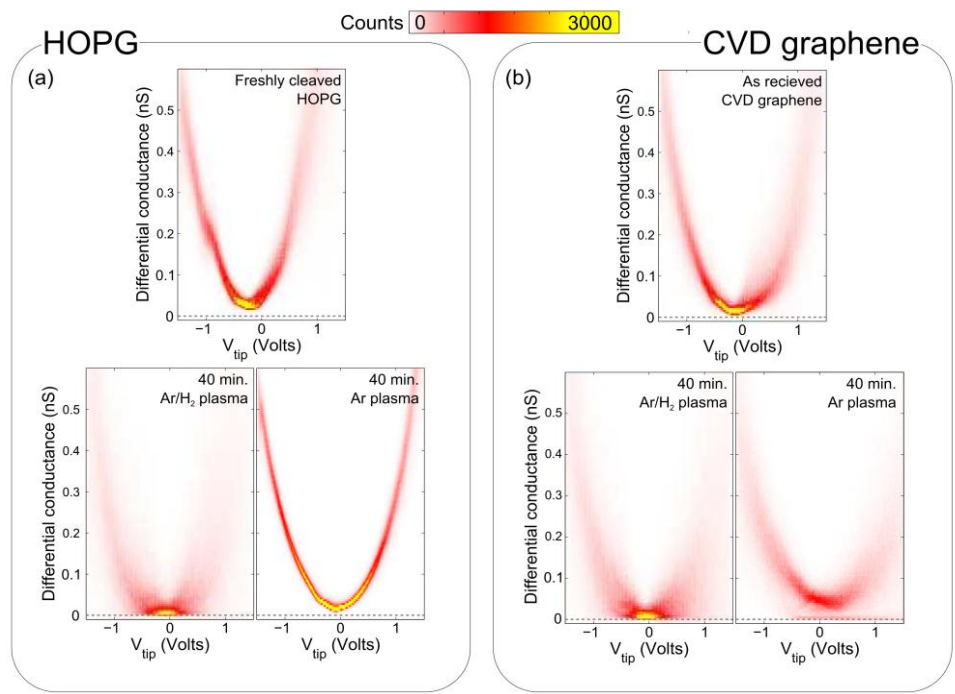

**Figure S1.** Two dimensional histograms of the differential conductance as a function of the tip bias voltage measured for pristine, $Ar/H_2$ plasma-treated and Ar plasma-treated samples: HOPG (a) and CVD graphene (b). The changes on their electronic properties are probed after the different treatments, measuring 2000 STS spectra at 50-100 different positions on the surface. The hydrogenation step is carried out by a 40 min. exposure to an $Ar/H_2$ plasma. The effect of Ar ion impacts on the electronic properties is studied by exposing the samples for 40 min. to an Ar plasma. For both the HOPG and CVD graphene samples, after the $Ar/H_2$ treatment, the 2D histograms show a semiconducting behavior (accumulation of data points around zero differential conductance value). Samples treated with the pure Ar plasma do not show this behavior.

While the samples treated with $Ar/H_2$ plasma present a high accumulation of traces with a flat region with zero differential conductance (indicating the opening of a bandgap), the samples treated with pure Ar plasma maintain a conducting behavior with only some residual traces with zero differential conductance which can be attributed to the presence of some adsorbed insulating molecules or lattice defects (see Figure S2).

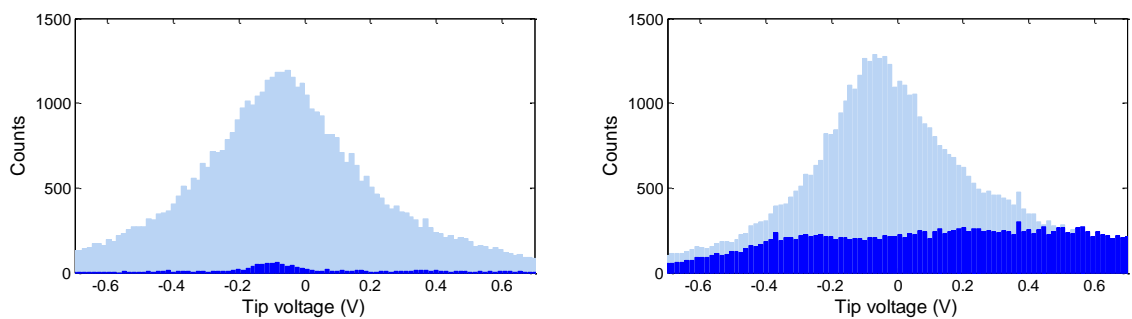

**Figure S2.** One dimensional histograms extracted from a profile along the dashed line (zero differential conductance) in the two dimensional histograms of Figure S1 for samples treated with the $Ar/H_2$ plasma (light blue) and Ar plasma (dark blue). Left panel corresponds to the HOPG and right panel to CVD graphene sample.

## 2D histograms of IV traces

The 2D histograms shown in Figure 2 of the manuscript are obtained by numerical differentiation of current *vs*. tip bias voltage traces (*IV* traces). Here we show the 2D histograms employing the *IV* traces without differentiation. From Figure S3, it is clear that the most probable *IV* traces for samples treated with the Ar/$H_2$ plasma show a flat region with zero current, indicating the opening of a bandgap around the Fermi level.

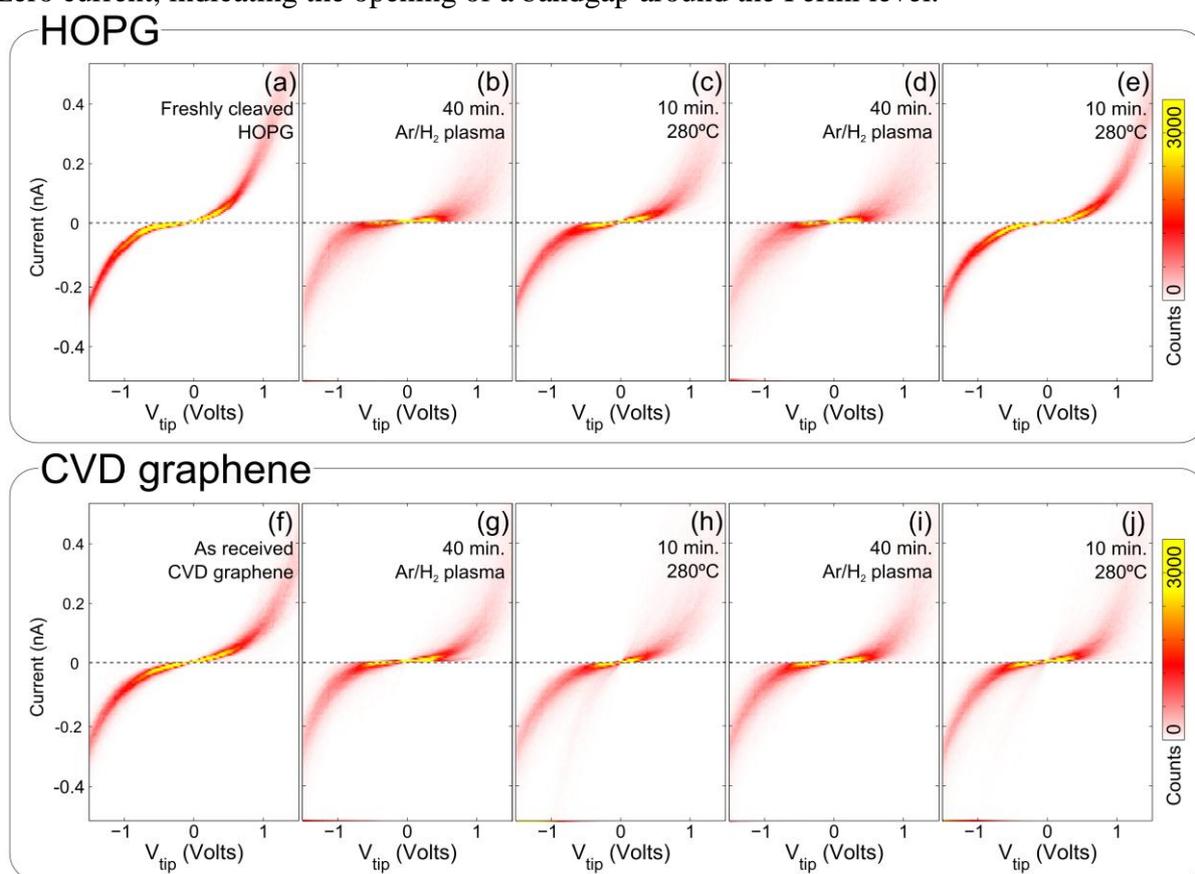

**Figure S3.** Two dimensional histograms of the tunneling current as a function of the tip bias voltage measured for HOPG (a-e) and CVD graphene (f-j) at different stages of the hydrogenation/dehydrogenation treatments. These 2D histograms are composed by 2000 of current *vs*. voltage traces measured at 50-100 different spots on the samples. The 2D histograms shown in Figure 2 of the manuscript have been prepared by numerical derivation of the IV traces employed in these histograms.

## 2$^{nd}$ hydrogenated CVD graphene sample

Here we show the results obtained for another graphene sample grown onto nickel. The main features shown in Figure 2 of the manuscript are also shown in this sample. After the plasma treatment, the sample shows a semiconducting behaviour which disappears after a moderate thermal annealing. Eventhough the higher heterogeneity of the CVD graphene samples the semiconducting behavior shown after Ar/$H_2$ plasma and the conducting behavior after annealing are robust results.

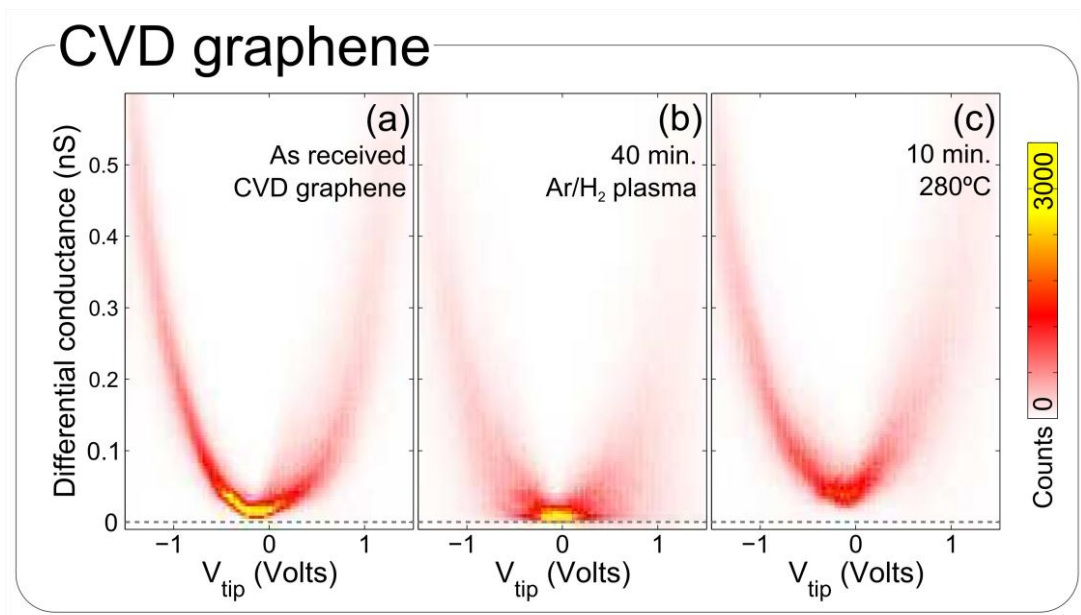

**Figure S4.** Two dimensional histograms of the differential conductance as a function of the tip bias voltage measured for another CVD graphene sample (different that the one used in the manuscript figures) at different steps of the hydrogenation/dehydrogenation process (a-c).